\documentclass[10pt,a4paper]{article}

\usepackage{amsmath,amssymb,amsthm}
\usepackage{url}
\usepackage{placeins}
\usepackage{subfigure}
\usepackage{multirow}
\usepackage{epsfig}
\usepackage{chicago}
\usepackage{rotating}

\newcommand\ben{\begin{equation}}
\newcommand\een{\end{equation}}
\newcommand\bea{\begin{eqnarray*}}
\newcommand\eea{\end{eqnarray*}}
\newcommand\bean{\begin{eqnarray}}
\newcommand\eean{\end{eqnarray}}

\begin{document}
\author{Andrew Green\footnote{Contact: andrew.green2@lloydsbanking.com}, Chris Kenyon\footnote{Contact: chris.kenyon@lloydsbanking.com} and Chris Dennis\footnote{Contact: chris.dennis2@lloydsbanking.com}}
\title{KVA: Capital Valuation Adjustment\footnote{\bf The views expressed are those of the authors only, no other representation should be attributed.}}
\date{First Submitted, February 24, 2014,\\
This Version, \today}

\maketitle

\begin{abstract}
Credit (CVA), Debit (DVA) and Funding Valuation Adjustments (FVA) are now familiar valuation adjustments made to the value of a portfolio of derivatives to account for credit risks and funding costs. However, recent changes in the regulatory regime and the increases in regulatory capital requirements has led many banks to include the \emph{cost of capital} in derivative pricing. This paper formalises the addition of cost of capital by extending the Burgard-Kjaer \citeyear{Burgard2013a} semi-replication approach to CVA and FVA to include an addition capital term, Capital Valuation Adjustment (KVA).\footnote{i.e. Kapital Valuation Adjustment to distinguish from CVA.} The utilization of the capital for funding purposes is also considered. The use of the semi-replication approach means that the flexibility around the treatment of self-default is carried over into this analysis. The paper further considers the practical calculation of KVA with reference to the Basel II \cite{BCBS-128} and Basel III \cite{BCBS-189} capital regimes and their implementation via CRD IV \cite{CRD-IV-Regulation,CRD-IV-Directive}. The paper also assesses how KVA may be hedged, given that any hedging transactions themselves lead to regulatory capital requirements and hence capital costs. Finally a number of numerical examples are presented to gauge the cost impact of KVA on vanilla derivative products.  
\end{abstract}
 
\section{Introduction}

Capital is a legal requirement for financial institutions holding derivatives, and requirements have increased over the past few years \cite{Dodd2010a,FED-2013-BIII,CRD-IV-Regulation,CRD-IV-Directive}. Hence it is surprising that few papers include capital in derivatives pricing, \cite{Kenyon2013d,Kenyon2014b,Hull2014a,Kenyon2014f}.  Here we extend the hedging framework of \cite{Burgard2011a,Kenyon2013a,Burgard2013a} to price capital requirements of derivatives trades by replicating its costs, together with the costs from credit and funding.  Thus we present a Capital Valuation Adjustment (KVA) alongside the existing adjustments for credit and funding. 

The Burgard-Kjaer semi-replication model has been selected because of the simplicity with which it can be extended to cover capital and because of the transparency of the treatment of cash flows. The PDE approach also makes it easy to explore the possible relationships between capital and both funding and default. 

Capital pricing appears challenging for several reasons:
\begin{enumerate}
	\item Diversity and length of regulations (see below).  In Basel III there are several major categories, each with hundreds of pages.
	\item Lifetime capital costs are needed, not just the spot costs.  
	\item Calculations must be done at several different levels of granularity and combined.  For example, for counterparty credit risk and CVA capital netting sets are important, while bank-level portfolio can also be needed such as for determination of the stressed period for Market Risk for SVAR calculation. Under the standardized approach CVA capital is calculated across all counterparties.\footnote{In practice for large numbers of counterparties it is well approximated by a summation over terms against individual counterparties as is described in section \ref{sec:CVAStd}.}
	\item The date when new regulation comes into force, and their exact content is often uncertain, for example a series of new regulations are currently in a consultation phase, \emph{fundamental review of the trading book} \cite{BCBS-219,BCBS-265}, (Revised) Standardized Approach \cite{BCBS-279}, \emph{margin requirements for non-centrally cleared derivatives} \cite{BCBS-261} and \emph{prudent valuation} \cite{EBA-CP-2013-28}.
\end{enumerate}
We present a brief list below of typical capital regulations here based on Basel III, and the type of calculation they require in Table \ref{t:types}.    

Whilst capital calculation may appear challenging, these calculations do not at first seem to introduce anything fundamentally different from CVA or FVA calculation. However, there are two truly new elements: firstly that hedging trades themselves generate capital requirements; and secondly the handling of capital itself. We include the capital requirements of hedges simply by always calculating the requirement of the entire portfolio. To deal with capital itself we introduce a parameter, $\phi$, to represent the fraction of the capital, $K$, used for funding. Capital used for funding represents the use of funds from issued equity capital. Clearly we have,
\begin{equation}
\phi\in [0,1].
\end{equation}
Funding with capital does reduce funding requirements, however, Basel III appears to explicitly prohibit linking capital issuance and its inverse to trading strategies.  Thus whilst a derivative can be funded by explicitly issuing-and-buying-back bonds, specific derivatives, or strategies, cannot be funded by issuing-and-buying back capital\footnote{
Derivatives require varying amounts of capital across their lifetimes. Under Basel III, we cannot trade in our own stocks (or own subordinated bonds) to meet this varying capital requirement because of Criterion 4 in CET1, Criterion 6 in Additional T1, and the 5Y-no-call in Tier 2, plus Criterion 5.b as below:
\begin{itemize}
	\item	Criterion 4  “The bank does nothing to create an expectation at issuance that the instrument will be bought back, redeemed or cancelled nor do the statutory or contractual terms provide any feature which might give rise to such an expectation”
\item	Criterion 6 “Any repayment of principal (eg through repurchase or redemption) must be with prior supervisory approval and banks should not assume or create market expectations that supervisory approval will be given”
\item	Criterion 5.b  “A bank must not do anything that creates an expectation that the call will be exercised”
\end{itemize}
}.  Allocating varying amounts of capital to derivatives over their lifetime is, of course, required as their capital requirements change.

In the numerical examples we consider two cases, the base-case $(\phi = 0)$, and  full use of capital for funding, $(\phi = 1)$. There are clearly many practical issues surrounding whether capital can be used to fund the derivative. To a degree this will be determined by the internal policy of the bank as to how capital is utilized. The base case $(\phi = 0)$ with no explicit use of capital may best reflect market practice. In addition, although capital may not be  used explicitly to fund derivatives its existence changes the funding requirements of the bank.  Thus $(\phi = 1)$ may be the most realistic case, although not reflected in current practice.  Hence current practice may not create full alignment between incentives and effects.  Of course, practically linking the cost of funding a trading desk with the capital available may be challenging.  We do not discuss the practicalities of different choices but leave this for further research. Systematic and theoretical consequences are dealt with elsewhere \cite{Kenyon2014b}.

\begin{table}
\noindent{
\begin{tabular}{|p{2cm}|p{3.6cm}|p{5.5cm}|}
\hline
 Classification & Alternatives & Calculation Type \\ \hline  
Counterparty & \emph{EAD Calculation}  & \\\cline{2-3}
 Credit Risk  & CEM & Function of Netting set Value \\
  & Standardized & Function of Netting set Value \\
	& Internal Model Method & Exposure profile\\\cline{2-3}
  & \emph{Weight Calculation} &  \\ \cline{2-3}
	& Standardized & External Ratings \\
	& FIRB & Internal \& External Ratings\\
	& AIRB & Internal \& External Ratings, Internal LGDs\\\hline 
	CVA Capital & Standardized &  Function of EAD\\ 
								& Advanced   &  VAR / SVAR on Regulatory CVA or CS01 \\ \hline
Market Risk  & Standardized & Deterministic formulae \\ 
& Internal Model Method &  VAR + SVAR  \\ \hline

\end{tabular}
}
\caption{Typical categories of capital regulations, their diversity (alternatives), and the type of calculations they require.} \label{t:types} \end{table}

The main contribution of this paper is to extend the pricing picture by including the costs of capital, the Capital Valuation Adjustment (KVA), in derivatives pricing by replication.  Given the increased regulatory focus on capital post-crisis, continuing regulatory developments, and its cost, this is long overdue.

\subsection{Diversity in Regulatory Capital Requirements}

Different quantities of capital are required for the same portfolio depending on the institution's regulatory status, and its interpretation of the regulations \cite{BCBS-267}.  The capital requirements also change depending on the intention of the institution. Hold-to-maturity positions (Banking Book) have different capital requirements to available-for-sale (Trading Book) \cite{BCBS-265}.  Our replication pricing is applicable to both these cases.  This does however mean that different institutions will have different replication costs, we go into detail on the implications of this in \cite{Kenyon2014b}.

In theory capital is a cost to risky businesses because investors require a positive return on risky investments.  We assert without proof that derivatives desks are risky businesses.  In practice capital use is charged by the issuing bank's treasury to derivatives desks.  This can be done by more, or less, direct methods, for example through budgets and RWA limits, but capital is always a cost to the desk.  

\section{Extending Semi-Replication to Include Capital}\label{sec:exsemi}

To include the cost of (regulatory) capital in pricing alongside Credit and Funding Valuation Adjustments we extend the semi-replication argument of Burgard and Kjaer \citeyear{Burgard2013a}. This paper uses the same notation as Burgard and Kjaer, table \ref{table:not} provides a summary. The sign convention is that the value of a cash amount is positive if received by the issuer. As with Burgard and Kjaer we seek to find the economic or shareholder value of the derivative portfolio, $\hat{V}$. Note also that here, as with Burgard and Kjaer \citeyear{Burgard2013a} we neglect balance sheet feedback effects. Burgard and Kjaer studied balance sheet feedback in \cite{Burgard2011b}. 

\begin{table}
\centering
\small
\begin{tabular}{|p{2cm}|p{10.5cm}|}\hline
{\bf Parameter} & {\bf Description}\\\hline 
$\hat{V}(t, S)$ & The economic value of the derivative or derivative portfolio\\
$V$ & The risk-free value of the derivative or derivative portfolio\\ 
$U$ & The valuation adjustment\\
$X$ & Collateral\\
$K$ & Capital Requirement\\
$\Pi$ & Replicating portfolio\\
$S$ & Underlying stock\\
$\mu_S$ & Stock drift\\
$\sigma_S$ & Stock volatility\\
$P_C$ & Counterparty Bond (zero recovery)\\
$P_1$ & Issuer bond with recovery $R_1$\\
$P_2$ & Issuer bond with recovery $R_2$, note $R_1\ne R_2$\\
$d\bar{\beta}_S$ & Growth in the cash account associated with stock (prior to rebalancing)\\
$d\bar{\beta}_C$ & Growth in the cash account associated with counterparty bond (prior to rebalancing)\\
$d\bar{\beta}_X$ & Growth in the cash account associated with collateral (prior to rebalancing)\\
$d\bar{\beta}_K$ & Growth in the cash account associated with capital (prior to rebalancing)\\
$r$ & Risk-free rate\\
$r_C$ & Yield on counterparty bond\\
$r_i$ & Yield on issuer bonds\\
$r_X$ & Yield on the collateral position\\
$r_F$ & Yield on issuer bond (one-bond case)\\
$M_B$ & Close-out value on issuer default\\
$M_C$ & Close-out value on counterparty default\\
$\alpha_C$ & Holding of counterparty bonds\\
$\alpha_i$ & Holding of issuer bond $i$\\
$\delta$ & The stock position\\
$\gamma_S$ & Stock dividend yield\\
$q_S$ & Stock repo rate\\
$q_C$ & Counterparty bond repo rate\\
$J_C$ & Default indicator for counterparty\\
$J_B$ & Default indicator for issuer\\
$g_B$ & Value of the derivative portfolio after issuer default\\
$g_C$ & Value of the derivative portfolio after counterparty default\\
$R_i$ & Recovery on issuer bond $i$\\
$R_C$ & Recovery on counterparty derivative portfolio\\ 
$\lambda_C$ & Effective financing rate of counterparty bond $\lambda_C= r_C - r$\\
$\lambda_B$ & Spread of a zero-recovery zero-coupon issuer bond. For bonds with recovery the following relation holds $(1 - R_i)\lambda_B = r_i - r$ for $i\in\{1,2\}$\\
$s_F$ & Funding spread in one bond case $s_F = r_F - r$\\
$s_X$ & Spread on collateral\\
$\gamma_K (t)$ & The cost of capital (the assets comprising the capital may themselves have a dividend yield and this can be incorporated into $\gamma_K (t)$)\\
$\Delta\hat{V}_B$ & Change in value of derivative on issuer default\\
$\Delta\hat{V}_C$ & Change in value of derivative on counterparty default\\
$\epsilon_h$ & Hedging error on default of issuer. Sometimes split into terms independent of and dependent on capital $\epsilon_h = \epsilon_{h_0} + \epsilon_{h_K}$\\
$P$ & $P = \alpha_1 P_1 + \alpha_2 P_2$ is the value of the own bond portfolio prior to default\\
$P_D$ & $P_D = \alpha_1 R_1 P_1 + \alpha_2 R_2 P_2$ is the value of the own bond portfolio after default\\
$\phi$ & Fraction of capital available for derivative funding\\\hline
\end{tabular}
\caption{\label{table:not}A summary of the notation, which is also common with Burgard and Kjaer (2013).}
\end{table}

The dynamics of the underlying assets are
\begin{align}
dS = & \mu_s S dt + \sigma_s S dW\\
dP_C = & r_C P_C dt - P_C dJ_C\\
dP_i = & r_i P_i dt - (1 - R_i)P_i dJ_B\quad\text{for}\quad i\in\{1,2\}
\end{align}
On default of the issuer, $B$, and the counterparty, $C$, the value of the derivative takes the following values
\begin{align}
\hat{V} (t, S, 1, 0) = & g_B(M_B, X)\\
\hat{V} (t, S, 0, 1) = & g_C(M_C, X).
\end{align}
The two $g$ functions allow a degree of flexibility to be included in the model around the value of the derivative after default. The usual assumption is that
\begin{align}
g_B = &(V - X)^+ + R_B(V-X)^- + X\nonumber\\
g_C = &R_C(V-X)^+ + (V-X)^- + X,
\end{align}
where
\begin{align}
x^+ = & \max\{x, 0\}\\
x^- = & \min\{x, 0\}.
\end{align}

 We assume that the following funding condition holds,
\begin{equation}\label{eq:bondfunding}
\hat{V} - X + \alpha_1 P_1 + \alpha_2 P_2 -\phi K = 0,
\end{equation}
where the addition of $\phi K$ represents the potential use of capital to offset funding requirements.
The growth in the cash account positions (prior to rebalancing, see Burgard and Kjaer \citeyear{Burgard2012b}; Brigo, Buescu, Pallavicini, and Liu \citeyear{Brigo2012a}) are given by,
\begin{align}
d\bar{\beta}_S = & \delta (\gamma_S - q_S) S dt\\
d\bar{\beta}_C = & -\alpha_C q_C P_C dt\\
d\bar{X} = & -r_X X dt.
\end{align}

In the portfolio, $\Pi$, we have to account for two different sources of regulatory capital requirements, the derivative and the replicating portfolio. Positions in the stock and counterparty bond will themselves attract a capital requirement. Hence we write that
\begin{equation}
K \equiv K(t, V, \text{``market risk"}, X, C, \delta, \alpha_C)
\end{equation}
reflecting the fact that the requlatory capital associated with the derivative is a function of the derivative portfolio value, its sensitivites through market risk capital, the collateral account value and the rating of the counterparty. The capital associated with the hedge portfolio is a function of the position in stock and bond. This effect reflects the fact that regulatory capital applies at the level of the whole derivative portfolio and not individual trades or counterparties. Some elements of the regulatory capital framework need to be attributed to portfolios from an overall net position. For example, market risk capital is calculated on the net position of all derivatives, while CVA capital under the standardized approach is calculated across all counterparties. 

The change in the cash account associated with the capital position is given by
\begin{equation}\label{eq:capcash}
d\bar{\beta}_K = -\gamma_K (t) K dt.
\end{equation}
This approach reflects the treatment of capital as a borrowing action, where capital is borrowed from shareholders to support derivative trading activities. The cost of capital is thus the cost of the return expected by shareholders for putting their capital at risk. In essence the derivatives business borrows the capital and pays cash profits to the shareholders at a given rate. It should also be noted that there is no term in $dJ_B$ and no impact on the default of the issuer. This reflects that any capital available to compensate the creditors of the issuer on default is already incorporated in the recovery rate $R_B$.\footnote{We have also considered the possibility that capital could be used to offset losses on counterparty default, which would lead to a term in $dJ_C$ in equation \eqref{eq:capcash}. There is an argument from symmetry to suggest that if capital is available for use in funding then it could also be used to offset losses on default. If we consider capital to be an exogenous resource and ignore balance sheet impact as is done in this paper this is appealing. However, ultimately we have rejected this on the grounds that it is unrealistic. Losses do directly impact the balance sheet and hence capital. To fully understand the interrelationship between counterparty default and capital requires a full balance sheet model.}  The final point to state is that we have implicitly assumed that the rating of the counterparty remains constant although the model could be extended to take account of rating transitions.

Using It\^{o}'s lemma the change in the value of the derivative portfolio is given by
\begin{equation}
d\hat{V} = \frac{\partial \hat{V}}{\partial t}dt + \frac{1}{2} \sigma^2 S^2 \frac{\partial^2\hat{V}}{\partial S^2}  dt +\frac{\partial \hat{V}}{\partial S} dS + \Delta \hat{V}_B dJ_B + \Delta \hat{V}_C dJ_C.
\end{equation}
Assuming the portfolio, $\Pi$, is self-financing, the change in its value is given by
\begin{equation}
\begin{split}
d\Pi = & \delta dS + \delta (\gamma_S - q_S) S dt + \alpha_1 dP_1 + \alpha_2 dP_2 + \alpha_C dP_C \\
&- \alpha_C q_C P_C dt - r_X X dt - \gamma_K K dt.
\end{split}
\end{equation}
Adding the derivative and replicating portfolio together we obtain
\begin{align}
d\hat{V} + d\Pi = & \Bigg[\frac{\partial \hat{V}}{\partial t} + \frac{1}{2} \sigma^2 S^2 \frac{\partial^2\hat{V}}{\partial S^2} + \delta (\gamma_S - q_S) S \nonumber\\
& + \alpha_1 r_1 P_1 + \alpha_2 r_2 P_2 + \alpha_C r_C P_C - \alpha_C q_C P_C - r_X X - \gamma_K K \Bigg] dt\\\nonumber
& + \epsilon_h dJ_B\\\nonumber
& + \left[\delta + \frac{\partial \hat{V}}{\partial S}\right] dS\\\nonumber
& + \left[g_C - \hat{V} - \alpha_C P_C\right] dJ_C,
\end{align}
where
\begin{align}\label{eq:ehdef}
\epsilon_h = & \left[\Delta\hat{V}_B - (P - P_D) \right]\\\nonumber
= & g_B - X + P_D - \phi K\\\nonumber
= & \epsilon_{h_0} + \epsilon_{h_K}
\end{align}
is the hedging error on the default of the issuer. In the final line the hedging error has been split into a term which does not depend on capital, $\epsilon_{h_0}$ and a term which does depend on capital $\epsilon_{h_K}$. In the event that $\phi=0$ then the hedge error is identical with that in Burgard and Kjaer. 

Assuming replication of the derivative by the hedging portfolio, except at the default of the issuer gives,
\begin{equation}
d\hat{V} + d\Pi = 0,
\end{equation}
and so make the usual assumptions to eliminate the remaining sources of risk so that
\begin{align}
\delta = & - \frac{\partial \hat{V}}{\partial S}\\
\alpha_C P_C = & g_C - \hat{V},
\end{align}
and this leads to the PDE
\begin{align}\label{eq:VhatPDE}
0 = & \frac{\partial \hat{V}}{\partial t} + \frac{1}{2} \sigma^2 S^2 \frac{\partial^2\hat{V}}{\partial S^2} - (\gamma_S - q_S) S \frac{\partial \hat{V}}{\partial S} - (r + \lambda_B + \lambda_C) \hat{V}\nonumber\\
& + g_C\lambda_C + g_B \lambda_B - \epsilon_h \lambda_B - s_X X - \gamma_K K + r \phi K\nonumber\\
& \hat{V}(T, S) = H(S).
\end{align}
where the bond funding equation \eqref{eq:bondfunding} has been used along with the yield of the issued bond, $r_i = r + (1- R_i) \lambda_B$\footnote{Note the this expression and this paper assumes zero bond-CDS basis.} and the definition of $\epsilon_h$ in equation \eqref{eq:ehdef} to derive the result,
\begin{equation}
\alpha_1 r_1 P_1 + \alpha_2 r_2 P_2 = rX -(r + \lambda_B)\hat{V} -\lambda_B (\epsilon_h - g_B) + r \phi K.
\end{equation}

Writing the derivative portfolio value, $\hat{V}$, as the sum of the risk-free derivative value, $V$ and a valuation adjustment $U$ and recognising that $V$ satisfies the Black-Scholes PDE,
\begin{align}
\frac{\partial V}{\partial t} + \frac{1}{2} \sigma^2 S^2 \frac{\partial^2 V}{\partial S^2} - (\gamma_S - q_S) S \frac{\partial V}{\partial S} - rV = & 0\nonumber\\
V(T, S) = & 0,
\end{align}
allows a PDE to be formed for the valuation adjustment,
\begin{align}
 \frac{\partial U}{\partial t} & + \frac{1}{2} \sigma^2 S^2 \frac{\partial^2 U}{\partial S^2} - (\gamma_S - q_S) S \frac{\partial U}{\partial S} - (r + \lambda_B + \lambda_C) U = \nonumber\\
& V \lambda_C - g_C \lambda_C + V \lambda_B - g_B \lambda_B + \epsilon_h \lambda_B + s_X X + \gamma_K K - r \phi K\nonumber\\
& U(T, S) = 0
\end{align}
Hence formally applying the Feynman-Kac theorem gives (using the terminology of Burgard and Kjaer),
\begin{equation}
U = \text{CVA} + \text{DVA} + \text{FCA} + \text{COLVA} + \text{KVA},
\end{equation}
where
\begin{align}
\text{CVA} = & -\int_t^T \lambda_C(u) e^{-\int_t^u (r(s) + \lambda_B(s) + \lambda_C(s)) ds}\nonumber\\
& \times \mathbb{E}_t \left[V(u) - g_C(V(u), X(u))\right] du\label{eq:intCVA}\\
\label{eq:intDVA}\text{DVA} = & -\int_t^T \lambda_B(u) e^{-\int_t^u (r(s) + \lambda_B(s) + \lambda_C(s)) ds}\mathbb{E}_t \left[V(u) - g_B(V(u), X(u))\right] du\\
\label{eq:intFCA}\text{FCA} = & -\int_t^T \lambda_B(u) e^{-\int_t^u (r(s) + \lambda_B(s) + \lambda_C(s)) ds}\mathbb{E}_t \left[\epsilon_{h_0}(u)  \right]du\\
\label{eq:intCOLVA}\text{COLVA} = & -\int_t^T s_X(u) e^{-\int_t^u (r(s) + \lambda_B(s) + \lambda_C(s)) ds} \mathbb{E}_t\left[X(u)\right] du\\
\label{eq:intKVA}\text{KVA} = & -\int_t^T e^{-\int_t^u (r(s) + \lambda_B(s) + \lambda_C(s)) ds} \nonumber\\
& \times \mathbb{E}_t \left[(\gamma_K (u) - r(u) \phi) K(u)+ \lambda_B \epsilon_{h_K}(u)\right] du.
\end{align}
In these expressions the FCA contains only the classical non-capital dependent hedging error, while the capital dependent terms have been grouped in KVA. Alternatively we could have grouped the additional term in the KVA integral in the FCA integral, to reflect the offset with funding, giving modified terms,
\begin{align}
\text{FCA}^{\prime} = & -\int_t^T (\lambda_B(u) \mathbb{E}_t \left[\epsilon_h(u)  \right]  - r(u) \phi \mathbb{E}_t \left[K(u)\right]) e^{-\int_t^u (r(s) + \lambda_B(s) + \lambda_C(s)) ds}du\\
\text{KVA}^{\prime} = & -\int_t^T \gamma_K (u) e^{-\int_t^u (r(s) + \lambda_B(s) + \lambda_C(s)) ds} \mathbb{E}_t \left[K(u)\right] du.
\end{align}

Whatever arrangement of the terms is selected, the capital elements resolve to calculating integrals over the capital profile $\mathbb{E}_t[K(u)]$ which is a strictly positive quantity. The generation of this profile is the subject of section \ref{sec:CalcKVA}.

\section{Capital at Portfolio Level}\label{sec:hedgingKVA}

Regulatory capital is a portfolio level requirement. The above model describes the calculation of KVA for an individual counterparty, while what we are actually interested in understanding is the total KVA for the whole portfolio, that is
\begin{equation}\label{eq:KVATot}
\text{KVA}_{\text{TOT}} = \sum_i^{\text{all ctpy's and hedge securities}} \text{KVA}_i,
\end{equation}
although in practice this may not be a simple sum. This is no great surprise as CVA and FVA desks, for example, in general manage the total CVA and FVA. Counterparty credit sensitivities may be hedged individually in some cases, particularly if the are appropriate single name CDS contracts available. However, the interest rate and other market risk of the CVA portfolio will be hedged across all counterparties. 

When pricing derivatives it is no longer sufficient to look at the impact of just the new trade, the impact of the trade and all hedging transactions should be considered. The hedge trades will themselves create additional capital requirements, although they may also mitigate other capital requirements. Consider a ten year interest rate swap traded with a corporate client on an unsecured basis. This trade has market risk, counterparty credit risk and CVA capital requirements associated with it. To hedge the market risk the trading desk enters another ten year swap with a market counterparty on a collateralised basis. This hedge trade generates a small amount of counterparty credit risk and CVA capital but drastically reduces the market risk capital on the whole book. 

KVA itself, like CVA and FVA, has market risk sensitivities. The Counterparty Credit Risk (CCR) term, for example, is clearly driven by the Exposure-at-Default (EAD) and hence by the exposure to the counterparty. Capital requirements go up as exposures rise irrespective to any impact on credit quality. KVA could be hedged and KVA hedging could be viewed as using trades to generate retained profits to offset additional capital requirements arising from market moves. However, KVA hedges will again generate capital requirements.  Although hedging trades generate capital requirements, because capital is generally a small percentage O(1/10), requirements converge quickly. Collateral will mitigate the additional CCR and CVA capital positions but the Market Risk capital will be affected as the hedge trade will look like a naked market risk position under the current capital regime.\footnote{Note that a similar situation has been avoided in the context of CVA capital for CDS spread hedges. Qualifying CDS positions that are designated as CVA hedges are exempt from further capital requirements.}   In the Numerical Examples we include Market Risk hedging and IR01 hedging for comparison with a naked position.

\subsection{Cost of Capital}
$\gamma_K(t)$ represents the cost of capital and its value has a direct impact on the size of the KVA adjustment. A key question is how should a value be assigned to it? The cost of capital represents the percentage return on regulatory capital that must be paid to shareholders in response to their decision to deploy this capital in support of derivatives trading activity. This means that the cost of capital is, in the view of the authors, an internal parameter that is set by the Banks's board of directors in consultation with shareholders. Hence in common with the cost of funding, cost of capital is idiosyncratic and may not be externally visible. A relatively close proxy is the return on equity target that is sometimes stated by individual banks. A recent report by Roland Berger and Nomura \cite{Reboul2014a} suggested that a typical group ROE target might be 10\%. However, the returns required by shareholders for individual banking activities may be higher or lower that the group level target. In the numerical examples in section \ref{sec:numexample} a value of 10\%\ has been used for $\gamma_K$.

\section{Calculating KVA for Regulatory Capital}\label{sec:CalcKVA}

In this section we consider the KVA associated with a derivative portfolio with a single counterparty. The aggregated KVA position will be obtained from equation \eqref{eq:KVATot} and as noted earlier this will require some capital attribution down to portfolio level. 

Here we will only consider the three main capital requirements that most derivative trades are subject to, Market Risk Capital, Counterparty Credit Risk Capital and Credit Valuation Adjustment Capital. Hence we can divide $K(u)$ up into three separate terms,
\begin{equation}
K = K_{\text{MR}} (u, \frac{\partial V}{\partial S}) + K_{\text{CCR}}(u, V, C, X) + K_{\text{CVA}} (u, V, C, X).
\end{equation}
Here the Market Risk is written as a function of the sensitivity of the unadjusted value $V$ to reflect the fact that it is driven by Market Risk, while the other terms are written as functions of the value, the collateral and of the properties of the counterparty.

\subsection{Market Risk Capital}
Market Risk Capital is a capital requirement held to offset against the risk of losses due to market risk on traded products and forms part of the Basel II framework \cite{BCBS-128}. As currently implemented market risk capital can be calculated in two ways, \emph{Standardised Method} and \emph{Internal Models Method (IMM)} for those institutions with appropriate regulatory approvals. Changes to the market risk capital framework are included in the \emph{Fundamental Review of the Trading Book} \cite{BCBS-265} but these changes will not be considered further here as the implementation date is unknown and the final proposals are not yet available.\footnote{The proposed changes would change certain aspects of the calculation under both Standardised and IMM approaches.}

It should be noted that in the case of both standardised and IMM approaches the market risk capital is calculated on a net basis across the portfolio. This is problematic from a calculation perspective as it implies that the market risk capital associated with a given portfolio will need to be attributed from the overall net requirement.  

One argument that could be made is that the market risk is hedged in full on a back-to-back basis and that in such circumstances the market risk capital is zero. If trades are hedged back-to-back then they can be taken out of the market-risk capital regime entirely. However, the existence of valuation adjustments means that in this case the overall delta will not be zero, even if the capital requirement is. A true hedge of portfolio delta, adjusting for the delta on valuation adjustments would give zero delta but still have a capital requirement as the valuation adjustments do not feature in the current market risk capital regime.\footnote{That is CVA capital is treated separately in the regime and is not part of the core framework as a valuation adjustment.}

To evaluate the impact of the the market risk on a net basis, section \ref{sec:numexample} explores both unhedged and hedged examples. For the hedged examples two cases are explored, one where an interest rate swap is hedge with an identical back-to-back transaction and one where the net portfolio IR01, including valuation adjustments, is reduced to zero at inception. 

\subsubsection{Standardised Method}
The standardised measurement method for Market Risk resolves to a formula based approach to generating the capital requirement with different approaches for interest rate, equity, foreign exchange and commodities risk. In each case there are a number of different optional approaches to the calculation available to the bank. Options are treated separately, again with multiple ways of quantifying the capital requirement. It is not the purpose of this paper to describe all of these approaches in detail and the reader is referred to the Basel II documentation \cite{BCBS-128} for a detailed description. However, the numerical examples in section \ref{sec:numexample} will be based on interest rate swaps and so the selected approach is summarised here.

An interest rate swap is treated as two positions in government securities, that is a notional position in a floating rate instrument with a maturity equal to the period until the next interest rate fixing and an opposite position in a fixed-rate instrument with a maturity equal to the residual maturity of the swap. Consider a GBP interest rate swap with a maturity of 10 years, where the bank pays a fixed rate of 2.7\% on a notional of GBP 100m, and receives 3 month LIBOR. The swap has an annual payment frequency and we assume that the first coupon has exactly three months to the next fixing. 

Assuming the use of the \emph{maturity method}, at the start of the simulation the floating leg will give a risk falling into the \emph{3 to 6 month} time-band and hence a risk weight of 0.40\%. The fixed leg will fall into the  \emph{9.3 to 10.6 years} time-band as the coupon is less than 3\%, giving a risk weight of 5.25\%. For a portfolio the short and long positions are summed in each band to give a weighted long and weighted short position. A \emph{vertical disallowance} equal to 10\% of the smaller of the weighted long and weighted short in each band would then be calculated. Banks then conduct two levels of \emph{horizontal offsetting} within three wider time zones spanning, 0-12 months, 1-5 years and 5+ years. Within each band there is a \emph{horizontal disallowance} and a second one between bands. The disallowance between zones 1 and 3 is 100\%. This then yields an overall market risk capital figure. A good description of the practical implementation of the standardized method is given in BIPRU \cite{fca:bipru}.

It should be clear from the above discussion that the market risk capital under standardized method for market risk is simply a function of trade properties such as residual maturity, coupon and notional. It is not a function of the current mark to market or risk. Hence the KVA formula reduces to
\begin{align}
\text{KVA}^{\text{std}}_{\text{MR}} = & -\int_t^T \gamma_K (u) e^{-\int_t^u (r(s) + \lambda_B(s) + \lambda_C(s)) ds} \mathbb{E}_t \left[K^{\text{std}}_{\text{MR}}(u)\right] du.\nonumber\\
= & -\int_t^T \gamma_K (u) e^{-\int_t^u (r(s) + \lambda_B(s) + \lambda_C(s)) ds} K^{\text{std}}_{\text{MR}}(u, M_i, S_i, N_i)du
\end{align}
where $M_i(u)$, $S_i(u)$ and $N_i(u)$ are the residual maturity, coupon and notional respectively for trade $i$. The inner expectation has dropped out and a Monte Carlo simulation is not required for this calculation. In practice this would be calculated using a simple numerical integral and this is done in the examples below.

\subsubsection{IMM}\label{sec:MRIMM}
The exact methodology used for internal model method market risk depends on an internal choice made by the bank in question and agreed with the appropriate regulatory body. The general approach is the same in all cases, however, using Value-at-Risk at the 99th percentile with price shocks generated from 10 day movements in prices.\footnote{Under the current proposals contained in the \emph{Fundamental Review of the Trading Book}, expected shortfall (CVAR) will replace VAR and the price shocks will be in most cases taken over periods longer than 10 days \cite{BCBS-265}.} The time-series of data must be at least a year. A number of different VAR methodologies can be used including variance-covariance, historical simulation and Monte Carlo. VAR models can also use full revaluation or \emph{delta-gamma-vega} approximation (that is, a Taylor series). 

All these IMM approaches to market risk capital will be expensive to compute $K_{\text{MR}}$ as they will typically involve Monte Carlo within Monte Carlo. So for example, a historical simulation full re-valuation model would require a historical simulation and full revaluation at each point inside the outer Monte Carlo that captures the capital exposure. This paper will not address the use of IMM for market risk, rather the reader is referred to Green and Kenyon \citeyear{Green2014a} for details of a suitable computational technique to accelerate this calculation.\footnote{Green and Kenyon \citeyear{Green2014a} examines the calculation of the cost of VAR-based initial margin, but this approach can be directly translated to the calculation of market risk capital.} 

\subsection{Counterparty Credit Risk Capital}

Counterparty Credit Risk Capital (CCR) is calculated for OTC derivatives using
\begin{equation}
{\rm RWA} = w\times 12.5\times {\rm EAD}
\end{equation}
where $w$ is the weight and ${\rm EAD}$ is the (regulatory) \emph{Exposure at Default} of the counterparty. The calculation methodology is divided into two separate parts to estimate the weight and the {\rm EAD}. The weight can be calculated using three different approaches in order of increasing sophistication and regulatory approval, \emph{Standardized Approach}, \emph{Foundation Internal Rating-Based (FIRB)} and \emph{Advanced Internal Rating-Based (AIRB)}. The ${\rm EAD}$ can be calculated using three different approaches, two simplified approaches based on trade mark-to-markets \emph{Current Exposure Method (CEM)} and \emph{Standardized} and the \emph{Internal Model Method (IMM)} using the banks own internal expected exposure engine.\footnote{CEM and Standardized will be replaced by the (Revised) Standardized Approach \cite{BCBS-279} at some point in the future.} 

\subsubsection{Weight Calculation}

\paragraph{Standardized Method}

In the Standardized Approach the weight is simply given by the external rating of the counterparty and the sector in which it operates. For unrated counterparties the weight is set at 100\%. Tables of the weights can be found in \cite{BCBS-128}.

\paragraph{Internal Ratings-Based Approach}

In the Internal Ratings-Based (IRB) approach banks estimate key risk components themselves: the probability of default (PD) and the loss given default (LGD). In the Foundation IRB approach banks provide PD estimates but use supervisory estimates for the LGD.  In the Advanced-IRB approach banks are also allowed to estimate the LGD. In both cases the weight is calculated according to the following formula,
\begin{align}
\rho =& 0.12 \frac{1-e^{-50\times {\rm PD}}}{1-e^{-50}} + 0.24\frac{1-(1-e^{-50\times {\rm PD}})}{1-e^{-50}} \\
b=& (0.11852 - 0.05478 \log({\rm PD}))^2 \\
w=& {\rm LGD}\left( \Phi\left( \frac{\Phi^{-1}({\rm PD})}{\sqrt{1-\rho}} + \Phi^{-1}(0.999)\sqrt{\frac{\rho}{1-\rho}}\right) - {\rm PD}\right)\\
&{}\times \frac{1+(M-2.5)b}{1-1.5b}
\end{align}
where $\Phi$ is the cumulative Normal distribution, and $\Phi^{-1}$ its inverse.

The PD is the greater of 0.03\%\ and the bank's internal estimate for probability of default over one year. Under FIRB the LGD = 45\%\ for corporates. $M$ is the effective maturity of the netting set and this is given by
\begin{equation}
M = \min\left(5.0, \max\left(1.0, \frac{\sum_{i=1}^{N_{\text{trades}}} m_i N_i}{\sum_{i=1}^{N_{\text{trades}}} N_i}\right)\right),
\end{equation}
where $m_i$ is the residual trade maturity and $N_i$ is the trade notional. 

\subsubsection{EAD Calculation}

\paragraph{EAD using CEM}

\newcommand\NGR{\ensuremath{{\rm N_{GR}}}}

In the CEM banks must get replacement costs by marking contracts to market, and then add a factor (the \emph{add-on}) to capture exposure over the remainder of the contract life. Hence the EAD is given by
\begin{equation}
{\rm EAD} = V + A(m_i, N_i, {\rm asset class}).
\end{equation} 
The add-on reflects the asset class (Interest Rates, FX and Gold, Equities, Other Precious Metals, Other Commodities) and the remaining maturity (less than one year, one to five, longer).  Add-ons are deterministic percentages of the contract notional.   

Some legally-supported bilateral netting is permitted with the net add-on $A_{\rm Net}$ calculated as:
\[
	A_{\rm Net} = 0.4 A_{\rm Gross} + 0.6 \NGR  A_{\rm Gross}
\]
where \NGR\ is the ratio of net to gross replacement costs and $A_{\rm Gross}$ is the gross add-on amount. The net to gross ratio is given by
\begin{equation}
{\text{NGR}} = \frac{(\sum_{i=1}^{N_{\text{trades}}} V_i)^+}{\sum_{i=1}^{N_{\text{trades}}} (V_i)^+}.
\end{equation}

For a single uncollateralized 10Y IR swap, the add-on for EAD is 1.5\%\ of notional. The CEM approach to calculating EAD will be adopted in the numerical examples.

\paragraph{EAD using Standardized Approach}  

\newcommand\CMV{\ensuremath{{\rm V_{transaction\, }}}}
\newcommand\CMC{\ensuremath{{\rm V_{collateral\, }}}}
\newcommand\RPT{\ensuremath{{\rm R_{transaction\, }}}}
\newcommand\RPC{\ensuremath{{\rm R_{collateral\, }}}}
\newcommand\CCF{\ensuremath{{\rm CCF}}}
\newcommand\EAD{\ensuremath{{\rm EAD}}}
\newcommand\CVA{\ensuremath{{\rm CVA}}}
\newcommand\EE{\ensuremath{{\rm EE}}}
\newcommand\D{\ensuremath{{\rm D}}}
\newcommand\LGD{\ensuremath{{\rm LGD}}}

For the Standardized Approach the EAD is calculated as:
\begin{equation}
\begin{split}
{\rm EAD} =& \beta\times\max\left(\sum_i \CMV_i - \sum_l \CMC_l, \right.    \\
& \qquad\qquad\ \ \left.  \sum_j \Big|  \sum_i \RPT_{ij} - \sum_l \RPC_{lj}    \Big| \times \CCF_j
\right)
\end{split}
\end{equation}
Where: \CMV\ value of transactions; \CMC\ value of collateral; \RPT\ risk from transactions; \RPC\ risk from collateral; \CCF\ supervisory credit conversion factor for the hedging set.  Indices: $i$ for transactions; $j$ for supervisory-designated hedging sets, these correspond to risk factors; $l$ collaterals.

For all transactions with a linear risk profile, apart from debt instruments and payment legs, the size of the risk position is the effective notional value of the underlying financial instrument converted to the firms reporting currency. For debt instruments and payment legs the risk is given by the effective notional value of the remaining payments multiplied by the modified duration and converted to the reporting currency. For an OTC derivative with a non-linear risk profile the risk position is given by the delta equivalent effective notional value of the underlying instrument except where that is a debt instrument. For OTC derivatives with a underlying debt instrument or payment leg the risk is the delta equivalent effective notional value of the financial instrument or payment leg multiplied by its modified duration. The risk on a collateral position is given by assuming that collateral received is a claim on the counterparty that is due today while posted collateral should be treated as an obligation to the counterparty due today.\footnote{See the description in Basel II \cite{BCBS-128} and BIPRU \cite{fca:bipru}.} 

The \CCF\ for debt positions are: $0.6\%$ when there is high specific risk; $0.3\%$ for a reference debt instrument beneath a CDS and has low specific risk; $0.2\%$ otherwise.  $\beta$ is set to $1.4$.

\paragraph{Calculating KVA for formula-based Approaches}
In both standardized method and CEM the EAD is a function of the value of the trades in the netting set $V_i$. Hence we calculate the KVA as follows
\begin{equation}\label{eq:KVAccrform}
\text{KVA}^{\text{formula}}_{\text{CCR}} = -\int_t^T \gamma_K (u) e^{-\int_t^u (r(s) + \lambda_B(s) + \lambda_C(s)) ds} 12.5 c\mathbb{E}_t \left[w EAD(V_i, u)\right] du,
\end{equation}
where $c$ is the capital multiplier that is currently set at 8\%. The inner expectation is broadly similar to the expected exposure calculation in CVA and FVA terms as the EAD at any point is simply a function of the portfolio value. 

\paragraph{\EAD\ under Internal Model Method}

\EAD\ is calculated according to the following formulae:
\begin{align*}
\EAD =& \alpha \times {\rm Effective\ EPE}  \\
{\rm Effective\ \EE_{t_k}} =& \max({\rm Effective\ \EE_{t_{k-1}}}, \EE_{k_i})\\
{\rm Effective\ EPE} =& \sum_{k=1}^{\min(1{\rm year,maturity})} {\rm Effective\ \EE_{t_k}}\times \Delta t_k\\
 \Delta t_k =& t_k - t_{k-1}\\
\alpha =& 1.4
\end{align*}
where \EE\ is the expected exposure (always greater than or equal to zero be definition of exposure). 

The formula for KVA has the same form as equation \eqref{eq:KVAccrform} with the inner expectation now given by
\begin{equation}
\mathbb{E}^Q \left[w \alpha \sum_{k=1}^{\min(1{\rm year,maturity})} \max({\rm Effective\ \EE(u)_{t_{k-1}}}, \EE(u)_{t_k}) \Delta t_k \Bigg|{\cal F}_t \right], 
\end{equation}
where\footnote{Note that in equation \eqref{eq:EEti} the measure is specified as the $Q$ measure. In practice the IMM approved model may in fact be set in the $P$ measure. This would add further complexity as the implied dynamics used to physically calculate the IMM exposures would be different from those used to estimate the KVA term which are risk-neutral. IMM is not restricted to use a $P$-measure exposure engine and the $Q$-measure is acceptable as long as the model fulfills the back-testing requirements.}
\begin{equation}\label{eq:EEti}
\EE_{t_k} = \mathbb{E}^Q\left[ \max(V(t_k), 0) |{\cal F}_{u} \right].
\end{equation}
The filtrations have been specified to aid clarity on exactly what expectations are being calculated. It is clear that we need to estimate future expected exposures inside the expectation used to give the EAD profile under IMM. This is problematic as it points to the need to use Monte Carlo within Monte Carlo to solve. American Monte Carlo is already widely used for such purposes by many practitioners \cite{Longstaff2001a,Cesari2010a}.

\subsection{CVA Capital}

CVA Capital was introduced in Basel III \cite{BCBS-189} in response to the large CVA losses some financial institutions faced during the 2007-2009 financial crisis. CRD-IV, the European implementation of Basel III, removes the requirement to calculate CVA Capital for corporate counterparties that are EU domiciled but it must still be calculated for other counterparties. Two methods of calculation are offered, \emph{standardized} and \emph{advanced} for those banks with IMM approval for both exposure and VAR calculation.

\subsubsection{Standardised} \label{sec:CVAStd}
The standardized CVA risk capital charge in \cite{BCBS-189}, paragraph 104, gives the formula to generate CVA capital:
\begin{align}
K_\CVA =&2.33\sqrt{h}\left\{ \left(\sum_i 0.5 \omega_i\left(M_i\EAD_i^{\rm total}-M_i^{\rm hedge}B_i\right) - \sum_{\rm ind} \omega_{\rm ind}M_{\rm ind}B_{\rm ind}\right)^2 \right. \nonumber \\
&
\left.\qquad\qquad\vphantom{\sum_{\rm ind}}   {}+ \sum_i 0.75 \omega_i^2\left(M_i\EAD_i^{\rm total}-M_i^{\rm hedge}B_i\right)^2  \right\}^{1/2} \label{e:standcva}
\end{align}
Where:
\begin{itemize}
\item $h$ one year risk horizon in units of years, i.e. h=1;
\item $\omega_i$ risk weight of $i^{\rm th}$ counterparty based on external rating (or equivalent);
\item $\EAD_i$ exposure at default of counterparty $i$, discounted using $\frac{1-e^{-0.05 M_i}}{0.05 M_i}$ (as we are using the non-IMM point of view);
\item $B_i$ notional of purchased single name CDS hedges, discounted as above;
\item $B_{\rm ind}$ notional of purchased index CDS hedges, discounted as above;
\item $\omega_{\rm ind}$ risk weight of index hedge using one of seven weights using the average index spread;
\item $M_i$ effective maturity of transactions with counterparty $i$, for non-IMM this is notional weighted average, and is not capped at five years;
\item $M_i^{\rm hedge}$ maturity of hedge instrument with notional $B_i$;
\item $M_{\rm ind}$ maturity of index hedge ${\rm ind}$.
\end{itemize}

The standardized CVA charge is calculated across all counterparties. Mitigation is given for CDS that are used to hedge counterparty credit risk. In the absence of hedging then formula reduces to
\begin{equation}
K_\CVA =2.33\sqrt{h}\left\{ \left(\sum_i 0.5 \omega_i M_i\EAD_i^{\rm total}\right)^2 + \sum_i 0.75 \omega_i^2 \left(M_i\EAD_i^{\rm total}\right)^2  \right\}^{1/2} \label{e:standcvanohedge}
\end{equation}
However, in the limit of a large number of counterparties it is well approximated as a sum over terms for individual counterparties,
\begin{equation}\label{eq:CVAstdLargeN}
K^i_\CVA \approx \frac{2.33}{2} \sqrt{h} \omega_i M_i \EAD_i^{\rm total}.
\end{equation}
This is a simple expression in terms of the EAD so KVA for CVA capital under the standardized approach for a single counterparty  is given by
\begin{equation}
\text{KVA}^{\text{std}}_{\text{CVA}} = -\int_t^T \gamma_K (u) e^{-\int_t^u (r(s) + \lambda_B(s) + \lambda_C(s)) ds} \mathbb{E}_t \left[\frac{2.33}{2} \sqrt{h} \omega M \EAD(V, u)\right] du.
\end{equation}

\subsubsection{IMM}

If the bank has Specific Interest Rate Risk VaR model approval and IMM approval for EAD calculation then it must use the Advanced CVA risk capital charge. The model uses VAR on the credit spread sensitivity of a unilateral CVA formula where the expected exposure is generated from a stressed calibration. Where the bank uses a VAR model with full revaluation then the following CVA formula must be used directly, 
\begin{align}
\CVA=&\LGD_{\rm MKT}\sum_{i=1}^T \max\left(0,\ \exp\left(-\frac{s_{i-1}t_{i-1}}{\LGD_{\rm MKT}} \right)  -  \exp\left(-\frac{s_{i}t_{i}}{\LGD_{\rm MKT}} \right)  \right)\nonumber\\
&\times  \left(\frac{\EE_{i-1}\D_{i-1}+\EE_{i}\D_{i}}{2}\right)
\end{align}
where \D\ are discount factors, and $s_i$ are market-observed CDS spreads. If the bank uses a VAR model based on credit spread sensitivities then the credit spread sensitivity is given by
\begin{equation}
\text{Regulatory CS01}_i = 0.0001 t_i \exp\left(-\frac{s_i t_i}{\LGD}\right)\left(\frac{\EE_{i-1}\D_{i-1}+\EE_{i+1}\D_{i+1}}{2}\right)
\end{equation}

The CVA capital under IMM is therefore given by the same approach as Market Risk Capital under IMM. To proceed we need to generate the forward expected exposures as was the case with CCR capital under IMM. As in that case, American Monte Carlo could be used to generate these. This would then need to be coupled with a \emph{lifetime} VAR technique as discussed in the context of IMM Market Risk Capital in section \ref{sec:MRIMM}.

\section{Numerical Examples}\label{sec:numexample}

Here we provide a number of example results to allow the impact of KVA to be assessed and compared to the existing valuation adjustments. In all cases the valuation adjustments have been calculated using numeric integration of equations \eqref{eq:intCVA} through \eqref{eq:intKVA}.  We choose to calculate the case of \emph{semi-replication with no shortfall at own default}, equivalent to ``strategy 1" in Burgard and Kjaer \citeyear{Burgard2013a}. Like Burgard and Kjaer we choose the first issuer bond to have zero recovery and use this bond to invest or fund the difference between $\hat{V}$ and V. The $P_2$ bond position has recovery $R_2 = R_B$ and is given by the funding constraint in equation \eqref{eq:bondfunding}. Hence we have,
\begin{align}
\alpha_1 P_1 = & - U\\
\alpha_2 P_2 = & -(V - \phi K).
\end{align}
Using these definitions of the hedge ratios gives the value of the issuer bond portfolio in default as
\begin{equation}
P_D = -R_B(V - \phi K)
\end{equation}
and hence $\epsilon_h = \epsilon_{h_0} + \epsilon_{h_K}$, is then given by
\begin{equation}
\epsilon_h = (1 - R_B) [V^+ - \phi K].
\end{equation}

This choice gives the following formulae for CVA, DVA, FCA and KVA for regular bilateral closeouts:
\begin{align}
\text{CVA} = & -(1-R_C)\int_t^T \lambda_C(u) e^{-\int_t^u(\lambda_B(s) + \lambda_C(s)) ds} \mathbb{E}_t \left[e^{-\int_t^u r(s)ds}(V(u))^+\right] du\label{eq:intCVAEx}\\
\label{eq:intDVAEx}\text{DVA} = & -(1-R_B)\int_t^T \lambda_B(u) e^{-\int_t^u (\lambda_B(s) + \lambda_C(s)) ds}\mathbb{E}_t \left[e^{-\int_t^u r(s)}ds (V(u))^-\right] du\\
\label{eq:intFCAEx}\text{FCA} = & -(1-R_B)\int_t^T \lambda_B(u) e^{-\int_t^u (\lambda_B(s) + \lambda_C(s)) ds}\mathbb{E}_t \left[e^{-\int_t^u r(s)ds}(V(u))^+ - \phi K \right]du\\
\label{eq:intKVAEx}\text{KVA} = & -\int_t^T e^{-\int_t^u (\lambda_B(s) + \lambda_C(s)) ds} \mathbb{E}_t \left[e^{-\int_t^u r(s)ds}K(u)(\gamma_K (u) - r(u) \phi)\right] du.
\end{align}
The funding term contains a term in capital and it is more convenient to group this in KVA,
\begin{align}
\text{FCA}^{\prime} = & -(1-R_B)\int_t^T \lambda_B(u) e^{-\int_t^u (\lambda_B(s) + \lambda_C(s)) ds}\mathbb{E}_t \left[e^{-\int_t^u r(s)ds}(V(u))^+  \right]du\\
\text{KVA}^{\prime} = & -\int_t^T e^{-\int_t^u (\lambda_B(s) + \lambda_C(s)) ds} \mathbb{E}_t \left[e^{-\int_t^u r(s)ds}K(u)(\gamma_K (u) - r_B(u) \phi)\right] du,
\end{align}
where we have used the fact that $r(u) + (1-R_B)\lambda_B = r_B(u)$. This second set of integrals shows that in the event $\phi$ is non-zero then the capital cost is reduced by the bank funding rate. 

Note that in these expressions, as we are examining an interest rate swap, interest rates are now assumed to follow a stochastic process so all short rates now appear inside  expectations.\footnote{The authors contend that the form of the XVA adjustments remain the same irrespective of the form of the dynamics for the underlying asset and remain the same even if interest rates are stochastic. Intuitively this can be seen from the form of the equations obtained and that they match what would be obtained for CVA and DVA using an expectation-based approach. It should also be noted that the Feynman-Kac theorem is general and applies to any form of It\^{o} process. A simple interest-rate model example is described in Appendix \ref{appendix:IR} and shows that the XVA adjustments remain the same even with stochastic rates.}

The examples have been calculated under the assumption the issuer calculates Market Risk under the standardized approach, uses the current exposure method to estimate the EAD and applies the standardized approach with external ratings for CCR and the standardized approach for CVA using the approximation for large numbers of counterparties give in equation \eqref{eq:CVAstdLargeN}. The use of the standardized approaches avoids the complexity and bespoke nature of internal model methods.  We assume that the issuer holds the minimum capital ratio requirement of 8\% (including minimum capital and capital buffer requirements) and that the issuer cost of capital, $\gamma_K$, is 10\%. 

The examples are calculated using a single 10 year GBP interest rate swap with semi-annual payment schedules. The fixed rate on the swap is 2.7\% ensuring the unadjusted value is zero at trade inception. We consider both the case where the issuer pays the fixed rate and the case where the issuer receives the fixed rate. The issuer spread information it is assumed to be flat 100bp accross all maturities and the issuer recovery rate is assumed to be 40\%.

We calculate all valuation adjustments for 4 different counterparty ratings and spread combinations, AAA, A, BB and CCC. The spreads assumed in each case are given in table \ref{tab:ctpyspread} alongside the risk-weight that is applied in the standardized CCR calculation. The counterparty recovery rate is assumed to be 40\%. 

\begin{table}[ht]
\centering
\begin{tabular}{|p{3cm}|r|p{3cm}|p{3cm}|}\hline
{\bf Counterparty Rating} & {\bf bp} & {\bf Standardized Risk Weight} &{\bf CVA Risk Weight $w_i$}\\\hline
AAA & 30 & 20\% & 0.7\%\\\hline
A & 75 & 50\% & 0.8\%\\\hline
BB & 250 & 100\% & 2\%\\\hline
CCC & 750 & 150\% & 10\%\\\hline
\end{tabular}
\caption{\label{tab:ctpyspread}Counterparty spread data used in the examples.}
\end{table}

\begin{table}
\centering
\resizebox{\columnwidth}{!}{%
\begin{tabular}{|c|l|l|rrr|rrr|r|r|}\hline
& & & & & & \multicolumn{3}{|c|}{\bf KVA} & & \\\hline
$\phi$ & {\bf Swap} & {\bf Rating} & {\bf CVA} & {\bf DVA} & {\bf FCA} & {\bf MR} & {\bf CCR} & {\bf CVA} & {\bf Total} & {\bf IR01} \\\hline
0 & \text{Pay} & \text{AAA} & -4 & 39 & -14 & -262 & -3 & -9 & -253.012 & 9.50816 \\
0 & \text{Pay} & \text{A} & -10 & 38 & -14 & -256 & -8 & -10 & -259.285 & 9.62228 \\
0 & \text{Pay} & \text{BB} & -31 & 33 & -12 & -234 & -14 & -22 & -279.175 & 10.0309 \\
0 & \text{Pay} & \text{CCC} & -68 & 24 & -9 & -185 & -16 & -87 & -341.55 & 11.2864 \\\hline
1 & \text{Pay} & \text{AAA} & -4 & 39 & -14 & -184 & -2 & -6 & -170.236 & 9.47109 \\
1 & \text{Pay} & \text{A} & -10 & 38 & -14 & -180 & -4 & -7 & -176.396 & 9.56193 \\
1 & \text{Pay} & \text{BB} & -31 & 33 & -12 & -166 & -7 & -16 & -198.05 & 9.90773 \\
1 & \text{Pay} & \text{CCC} & -68 & 24 & -9 & -134 & -8 & -63 & -259.724 & 10.9702 \\\hline
0 & \text{Rec} & \text{AAA} & -12 & 14 & -39 & -262 & -7 & -18 & -324.978 & -9.60701 \\
0 & \text{Rec} & \text{A} & -29 & 14 & -38 & -256 & -18 & -20 & -347.152 & -9.80112 \\
0 & \text{Rec} & \text{BB} & -84 & 12 & -33 & -234 & -31 & -46 & -416.404 & -10.4739 \\
0 & \text{Rec} & \text{CCC} & -177 & 9 & -24 & -185 & -34 & -176 & -587.071 & -12.3688 \\\hline
1 & \text{Rec} & \text{AAA} & -12 & 14 & -39 & -184 & -4 & -12 & -236.768 & -9.54678 \\
1 & \text{Rec} & \text{A} & -29 & 14 & -38 & -180 & -9 & -14 & -255.629 & -9.7039 \\
1 & \text{Rec} & \text{BB} & -84 & 12 & -33 & -166 & -16 & -32 & -318.491 & -10.2777 \\
1 & \text{Rec} & \text{CCC} & -177 & 9 & -24 & -134 & -18 & -123 & -467.039 & -11.8776 \\\hline

\end{tabular}%
}
\caption{\label{tab:KVAresults}XVA values for a GBP 10 year payers and 10 year receivers interest rate swap. Results are quoted in bp of the trade notional. The first column, $\phi$ specifies the use of capital for funding.}
\end{table}

The results of the example calculations are given in table \ref{tab:KVAresults}. Setting aside the Market Risk component of the capital we see that KVA from CCR and CVA terms gives an adjustment of similar magnitude to the existing CVA, DVA and FCA terms, demonstrating that KVA is a significant contributor to the price of the derivative. 

The market risk is assumed to be unhedged and so this KVA component is relatively large compared to the CCR and CVA terms. Under the standardized approach to market risk the capital requirement on a ten year transaction of this type is scaled according to a 60 bp move in rates. Practical applications would calculate the market risk capital requirement over all trades in a portfolio including hedges and then attribute these to trade level. The cases where the parameter $\phi$ is non zero show a reduction in capital costs. 

To assess the impact of hedging the second example consists of two interest rate swaps traded on a back-to-back basis so that the hedge trade is the exact mirror of the primary trade. For the hedge trade we assume perfect collateralisation so there is no CVA, DVA or FCA and that the collateral rate is equal to the risk free rate so that the COLVA is also zero. The KVA for the second trade is not zero even with perfect collateralisation. The market risk capital will be zero as the trades match exactly and hence can be removed from market risk capital under the Basel regulatory framework. The results for the combined portfolio are given in table \ref{table:KVAb2b}. In spite of the fact that the portfolio has no market risk capital it does have an open market risk position that comes from the valuation adjustment terms and the portfolio IR01 is given in the final column of the table. 

\begin{table}
\centering
\resizebox{\columnwidth}{!}{%
\begin{tabular}{|c|l|l|rrr|rrr|r|r|}\hline
& & & & & & \multicolumn{3}{|c|}{\bf KVA} & & \\\hline
$\phi$ & {\bf Swap} & {\bf Rating} & {\bf CVA} & {\bf DVA} & {\bf FCA} & {\bf MR} & {\bf CCR} & {\bf CVA} & {\bf Total} & {\bf IR01} \\\hline
0 & \text{Pay} & \text{AAA} & -4 & 39 & -14 & 0 & -3 & -9 & 9.2752 & 0.608158 \\
0 & \text{Pay} & \text{A} & -10 & 38 & -14 & 0 & -8 & -10 & -3.17873 & 0.722276 \\
0 & \text{Pay} & \text{BB} & -31 & 33 & -12 & 0 & -14 & -22 & -45.1125 & 1.1309 \\
0 & \text{Pay} & \text{CCC} & -68 & 24 & -9 & 0 & -16 & -87 & -156.464 & 2.38643 \\\hline
1 & \text{Pay} & \text{AAA} & -4 & 39 & -14 & 0 & -2 & -6 & 13.2661 & 0.57109 \\
1 & \text{Pay} & \text{A} & -10 & 38 & -14 & 0 & -4 & -7 & 3.22704 & 0.661926 \\
1 & \text{Pay} & \text{BB} & -31 & 33 & -12 & 0 & -7 & -16 & -32.3335 & 1.00773 \\
1 & \text{Pay} & \text{CCC} & -68 & 24 & -9 & 0 & -8 & -63 & -125.38 & 2.0702 \\\hline
0 & \text{Rec} & \text{AAA} & -12 & 14 & -39 & 0 & -7 & -18 & -62.6911 & -0.707008 \\
0 & \text{Rec} & \text{A} & -29 & 14 & -38 & 0 & -18 & -20 & -91.0454 & -0.901123 \\
0 & \text{Rec} & \text{BB} & -84 & 12 & -33 & 0 & -31 & -46 & -182.341 & -1.57391 \\
0 & \text{Rec} & \text{CCC} & -177 & 9 & -24 & 0 & -34 & -176 & -401.984 & -3.46885 \\\hline
1 & \text{Rec} & \text{AAA} & -12 & 14 & -39 & 0 & -4 & -12 & -53.2656 & -0.646782 \\
1 & \text{Rec} & \text{A} & -29 & 14 & -38 & 0 & -9 & -14 & -76.0064 & -0.803898 \\
1 & \text{Rec} & \text{BB} & -84 & 12 & -33 & 0 & -16 & -32 & -152.774 & -1.37766 \\
1 & \text{Rec} & \text{CCC} & -177 & 9 & -24 & 0 & -18 & -123 & -332.695 & -2.97758 \\\hline
\end{tabular}%
}
\caption{\label{table:KVAb2b}XVA values for a GBP 10 year payers and 10 year receivers interest rate swap with an identical back-to-back hedge. The interest rate swap hedge is assumed to be a counterparty that trades under a perfect CSA with instantaneous transfer of collateral, zero threshold and zero minimum transfer amount.  Results are quoted in bp of the trade notional. The market risk capital is now zero as the trade and hedge perfectly offset each other. The first column, $\phi$, specifies the use of capital for funding.}
\end{table}

If instead of using a back to back hedge the net portfolio market risk was eliminated at trade inception with a static hedge, the portfolio would have an IR01 of zero at the start. However, it would still attract market risk capital as the trade and hedge would not match exactly. This case is illustrated in table \ref{table:KVAhedge}. Again for the hedge trade we assume perfect collateralisation.  

The impact of allowing capital to be used as funding, $\phi=1$, is significant. KVA from the CCR term is reduced by approximately half, while that from CVA is reduced by around one third. Where the market risk capital is non-zero setting $\phi=1$ also reduces the KVA associated with it by approximately one third. 

\begin{table}
\centering
\resizebox{\columnwidth}{!}{%
\begin{tabular}{|c|l|l|rrr|rrr|r|r|p{1.2cm}|}\hline
& & & & & & \multicolumn{3}{|c|}{\bf KVA} & & &\\\hline
$\phi$ & {\bf Swap} & {\bf Rating} & {\bf CVA} & {\bf DVA} & {\bf FCA} & {\bf MR} & {\bf CCR} & {\bf CVA} & {\bf Total} & {\bf IR01} &{\bf Hedge Change ($\%$)} \\\hline
0 & \text{Pay} & \text{AAA} & -4 & 39 & -14 & -17 & -4 & -12 & -13 & 0 & 7 \\
0 & \text{Pay} & \text{A} & -10 & 38 & -14 & -20 & -11 & -13 & -30 & 0 & 8 \\
0 & \text{Pay} & \text{BB} & -31 & 33 & -12 & -28 & -20 & -31 & -88 & 0 & 12 \\
0 & \text{Pay} & \text{CCC} & -68 & 24 & -9 & -45 & -22 & -127 & -249 & 0 & 24 \\\hline
1 & \text{Pay} & \text{AAA} & -4 & 39 & -14 & -12 & -3 & -8 & -1 & 0 & 6 \\
1 & \text{Pay} & \text{A} & -10 & 38 & -14 & -13 & -5 & -9 & -14 & 0 & 7 \\
1 & \text{Pay} & \text{BB} & -31 & 33 & -12 & -18 & -9 & -23 & -59 & 0 & 11 \\
1 & \text{Pay} & \text{CCC} & -68 & 24 & -9 & -29 & -12 & -92 & -187 & 0 & 21 \\\hline
0 & \text{Rec} & \text{AAA} & -12 & 14 & -39 & -20 & -8 & -21 & -87 & 0 & 8 \\
0 & \text{Rec} & \text{A} & -29 & 14 & -38 & -25 & -20 & -23 & -122 & 0 & 10 \\
0 & \text{Rec} & \text{BB} & -84 & 12 & -33 & -40 & -36 & -54 & -234 & 0 & 17 \\
0 & \text{Rec} & \text{CCC} & -177 & 9 & -24 & -67 & -41 & -213 & -512 & 0 & 36 \\\hline
1 & \text{Rec} & \text{AAA} & -12 & 14 & -39 & -13 & -4 & -14 & -69 & 0 & 7 \\
1 & \text{Rec} & \text{A} & -29 & 14 & -38 & -16 & -10 & -16 & -95 & 0 & 9 \\
1 & \text{Rec} & \text{BB} & -84 & 12 & -33 & -25 & -18 & -38 & -186 & 0 & 15 \\
1 & \text{Rec} & \text{CCC} & -177 & 9 & -24 & -42 & -21 & -151 & -405 & 0 & 31 \\\hline
\end{tabular}%
}
\caption{\label{table:KVAhedge}XVA values for a GBP 10 year payers and 10 year receivers interest rate swap with a hedge adjusted to offset the portfolio IR01. The interest rate swap hedge is assumed to be a counterparty that trades under a perfect CSA with instantaneous transfer of collateral, zero threshold and zero minimum transfer amount.  Results are quoted in bp of the trade notional. The residual IR01 is now zero but market risk capital is non-zero as the trade and hedge do not perfectly offset each other from a capital perspective. The last column gives the adjustment to the hedge trade notional required to obtain a IR01 of zero. The first column, $\phi$, specifies the use of capital for funding.}
\end{table}

\section{Conclusions}\label{sec:conclusions}
We have presented a unified model for valuation adjustments that includes the impact of Capital and in so doing have introduced a new ``XVA" term, KVA. The impact of capital on funding has also been explored. We have described how KVA can be calculated in the case of the formula-based approached to regulatory capital calculation and sketched how this may be calculated in the case of internal model approaches. Practical examples of KVA on an interest rate swap have demonstrated how significant capital costs are, and that KVA is broadly similar in size to the other components of XVA. The use of capital, to reduce funding requirements ($\phi=1$) results in reductions in KVA of around one third to one half. However, it is not clear in practice if this option will be available to a derivatives trading desk.  In as much as this reflects a divergence of practice from actual effects some reassessment may be required.

The implication of the introduction of KVA is that just like CVA and FVA, KVA should be managed and hedged. KVA can be aligned with the counterparty and clearly has contingency on the survival of the counterparty and issuer. The most appropriate approach would be to manage KVA alongside CVA and FVA at portfolio level. KVA and capital management become part of the responsibility of a central \emph{resource management} desk.

\section*{Acknowledgements}
The authors would like to thank Lincoln Hannah, Glen Rayner and the two anonymous referees for their useful comments and contribution to the paper. 

\appendix
\section{XVA under Stochastic Interest Rates: A simple example}\label{appendix:IR}
To illustrate that the form of the XVA adjustment remains the same under a simple interest rate model consider the following example. In this case the derivative is a function of the value of a risk-free zero-coupon bond, $P_r(t, H)$, with dynamics
\begin{equation}
\frac{dP_r(t, H)}{P_r(t, H)} = r(t) dt + \sigma_r(t, H) dW,
\end{equation}
where $H > T$. We also assume that the counterparty and issuer bonds are also stochastic and that they are driven by the same Brownian motion alongside the jump to default process described above,
\begin{align}
\frac{dP_C(t, T)}{P_C(t, T)} =& r_C(t) dt + \sigma_C(t, T) dW -dJ_C\\
\frac{dP_i(t, T)}{P_i(t, T)} =& r_i(t) dt + \sigma_i(t, T) dW -(1-R_i) dJ_B.\\
\end{align}
The close-out conditions and funding equation remain as described above as do those for $d\bar{\beta}_C$, $dX$ and $d\bar{\beta}_K$. The equation for $d\bar{\beta}_S$ is now replaced with one for the holding, $\delta$ of risk-free bonds,
\begin{equation}
d\bar{\beta}_r = -\delta r P_r dt,
\end{equation}
which follows from the fact that the zero-coupon bond has no dividend yield and we assume that we can repo the bond at the risk-free rate. By It\^{o}'s lemma we obtain,
\begin{equation}
d\hat{V} = \frac{\partial \hat{V}}{\partial t}dt + \frac{1}{2} \sigma^2 P_r^2 \frac{\partial^2\hat{V}}{\partial P_r^2}  dt +\frac{\partial \hat{V}}{\partial P_r} dP + \Delta \hat{V}_B dJ_B + \Delta \hat{V}_C dJ_C.
\end{equation}
The change in the portfolio is given by
\begin{equation}
\begin{split}
d\Pi = & \delta dP_r - \delta r P_r  dt + \alpha_1 dP_1 + \alpha_2 dP_2 + \alpha_C dP_C \\
&- \alpha_C q_C P_C dt - r_X X dt - \gamma_K K dt + \Delta_K dJ_C.
\end{split}
\end{equation}
Combining these expressions gives,
\begin{align}
d\hat{V} + d\Pi = & \Bigg[\frac{\partial \hat{V}}{\partial t} + \frac{1}{2} \sigma^2 P_r^2 \frac{\partial^2\hat{V}}{\partial P_r^2} - \delta r P_r  + \alpha_1 r_1 P_1 + \alpha_2 r_2 P_2\\\nonumber
& + \alpha_C r_C P_C - \alpha_C q_C P_C - r_X X - \gamma_K K + r P_r \left(\frac{\partial \hat{V}}{\partial P_r} + \delta\right)\Bigg] dt\\
& + \epsilon_h dJ_B\\
& + \left[\delta + \frac{\partial \hat{V}}{\partial P_r}\right] dW + \alpha_C \sigma_C P_C dW + \alpha_1 \sigma_1 P_1 dW + \alpha_2 \sigma_2 P_2 dW\\
& + \left[g_C + \Delta_K - \hat{V} - \alpha_C P_C\right] dJ_C. 
\end{align}
In this case the form of delta to eliminate all the risk from the Brownian terms is given by,
\begin{equation}
\delta = -\frac{\partial\hat{V}}{\partial P_r} - \alpha_C\sigma_C P_C - \alpha_1\sigma_1 P_1 - \alpha_2\sigma_2 P_2,
\end{equation}
while the choice of $\alpha_C$ remains unchanged. The PDE for $\hat{V}$ has a very similar form to the earlier case for the stock underlying,
\begin{align}
0 = & \frac{\partial \hat{V}}{\partial t} + \frac{1}{2} \sigma_r^2 P_r^2 \frac{\partial^2\hat{V}}{\partial P_r^2} + r P_r \frac{\partial \hat{V}}{\partial P_r} - (r + \lambda_B + \lambda_C) \hat{V}\nonumber\\
& + (g_C + \Delta_K)\lambda_C + g_B \lambda_B - \epsilon_h \lambda_B - s_X X - \gamma_K K + r \phi K\nonumber\\
& \hat{V}(T, P_r(t, H)) = H(P_r),
\end{align}
where the bond funding equation and the expression for $\alpha_C$ have been used. Given this has the same form as equation \eqref{eq:VhatPDE} the subsequent analysis and results are the same, except that when the Feynman-Kac expressions for XVA are obtained the short rate must remain inside the expectation.

\bibliographystyle{chicago}
\bibliography{kenyon_general}

\begin{thebibliography}{}

\bibitem[\protect\citeauthoryear{BCBS-128}{BCBS-128}{2006}]{BCBS-128}
BCBS-128 (2006, {June}).
\newblock {International Convergence of Capital Measurement and Capital
  Standards}.
\newblock {\em Basel Committee for Bank Supervision\/}.

\bibitem[\protect\citeauthoryear{BCBS-189}{BCBS-189}{2011}]{BCBS-189}
BCBS-189 (2011).
\newblock {Basel III: A global regulatory framework for more resilient banks
  and banking systems}.
\newblock {\em Basel Committee for Bank Supervision\/}.

\bibitem[\protect\citeauthoryear{BCBS-219}{BCBS-219}{2012}]{BCBS-219}
BCBS-219 (2012).
\newblock {Fundamental review of the trading book --- consultative document}.
\newblock {\em Basel Committee for Bank Supervision\/}.

\bibitem[\protect\citeauthoryear{BCBS-261}{BCBS-261}{2013}]{BCBS-261}
BCBS-261 (2013).
\newblock {Margin requirements for non-centrally cleared derivatives}.
\newblock {\em Basel Committee for Bank Supervision\/}.

\bibitem[\protect\citeauthoryear{BCBS-265}{BCBS-265}{2013}]{BCBS-265}
BCBS-265 (2013).
\newblock {Fundamental review of the trading book - second consultative
  document}.
\newblock {\em Basel Committee for Bank Supervision\/}.

\bibitem[\protect\citeauthoryear{{BCBS-267}}{{BCBS-267}}{2013}]{BCBS-267}
{BCBS-267} (2013).
\newblock {Second report on the regulatory consistency of risk-weighted assets
  in the trading book issued by the Basel Committee}.
\newblock {\em Basel Committee for Bank Supervision\/}.

\bibitem[\protect\citeauthoryear{{BCBS-279}}{{BCBS-279}}{2014}]{BCBS-279}
{BCBS-279} (2014).
\newblock {The standardised approach for measuring counterparty credit risk
  exposures}.
\newblock {\em Basel Committee for Bank Supervision\/}.

\bibitem[\protect\citeauthoryear{Brigo, Buescu, Pallavicini, and Liu}{Brigo
  et~al.}{2012}]{Brigo2012a}
Brigo, D., C.~Buescu, A.~Pallavicini, and Q.~Liu (2012).
\newblock Illustrating a problem in the self-financing condition in two
  2010-2011 papers on funding, collateral and discounting.
\newblock {\em SSRN\/}.
\newblock http://ssrn.com/abstract=2103121.

\bibitem[\protect\citeauthoryear{Burgard and Kjaer}{Burgard and
  Kjaer}{2011a}]{Burgard2011b}
Burgard, C. and M.~Kjaer (2011a).
\newblock In the balance.
\newblock {\em {Risk}\/}~{\em 24\/}(11).

\bibitem[\protect\citeauthoryear{Burgard and Kjaer}{Burgard and
  Kjaer}{2011b}]{Burgard2011a}
Burgard, C. and M.~Kjaer (2011b).
\newblock Partial differential equation representations of derivatives with
  bilateral counterparty risk and funding costs.
\newblock {\em The Journal of Credit Risk\/}~{\em 7}, 75--93.

\bibitem[\protect\citeauthoryear{Burgard and Kjaer}{Burgard and
  Kjaer}{2012}]{Burgard2012b}
Burgard, C. and M.~Kjaer (2012).
\newblock {Addendum to: 'PDE Representations of Derivatives with Bilateral
  Counterparty Risk and Funding Costs'}.
\newblock {\em SSRN\/}.
\newblock Available at http://ssrn.com/abstract=2109723.

\bibitem[\protect\citeauthoryear{Burgard and Kjaer}{Burgard and
  Kjaer}{2013}]{Burgard2013a}
Burgard, C. and M.~Kjaer (2013).
\newblock {Funding Strategies, Funding Costs}.
\newblock {\em {Risk}\/}~{\em 26\/}(12).

\bibitem[\protect\citeauthoryear{Cesari, Aquilina, Charpillon, Filipovic, Lee,
  and Manda}{Cesari et~al.}{2010}]{Cesari2010a}
Cesari, G., J.~Aquilina, N.~Charpillon, Z.~Filipovic, G.~Lee, and I.~Manda
  (2010).
\newblock {\em {Modelling, Pricing, and Hedging Counterparty Credit Exposure: A
  Technical Guide }}.
\newblock London: Springer Finance.

\bibitem[\protect\citeauthoryear{{Department of the Treasury}}{{Department of
  the Treasury}}{2013}]{FED-2013-BIII}
{Department of the Treasury} (2013).
\newblock {12 CFR Parts 208, 217, and 225. Regulatory Capital Rules: Regulatory
  Capital Rules: Regulatory Capital, Implementation of Basel III, Capital
  Adequacy, Transition Provisions, Prompt Corrective Action, Standardized
  Approach for Risk-weighted Assets, Market Discipline and Disclosure
  Requirements, Advanced Approaches Risk-Based Capital Rule, and Market Risk
  Capital Rule; Final Rule}.
\newblock Federal Register, Vol. 78(198), pp62017-62291.
\newblock Department of the Treasury.

\bibitem[\protect\citeauthoryear{Dodd and Frank}{Dodd and
  Frank}{2010}]{Dodd2010a}
Dodd, C. and B.~Frank (2010).
\newblock {Dodd-Frank Wall Street Reform and Consumer Protection Act}.
\newblock H.R. 4173,
  \url{http://www.sec.gov/about/laws/wallstreetreform-cpa.pdf}.

\bibitem[\protect\citeauthoryear{EBA}{EBA}{2013}]{EBA-CP-2013-28}
EBA (2013).
\newblock {On prudent valuation under Article 105(14) of Regulation (EU)
  575/2013}.
\newblock Technical report, {European Banking Authority}.
\newblock EBA-CP-2013-28.

\bibitem[\protect\citeauthoryear{{EU}}{{EU}}{2013a}]{CRD-IV-Directive}
{EU} (2013a).
\newblock {Directive 2013/36/EU of the European Parliament and of the Council
  of 26 June 2013 on access to the activity of credit institutions and the
  prudential supervision of credit institutions and investment firms, amending
  Directive 2002/87/EC and repealing Directives 2006/48/EC and 2006/49/EC Text
  with EEA relevance}.
\newblock {\em European Commission\/}.

\bibitem[\protect\citeauthoryear{{EU}}{{EU}}{2013b}]{CRD-IV-Regulation}
{EU} (2013b).
\newblock {Regulation (EU) No 575/2013 of the European Parliament and of the
  Council of 26 June 2013 on prudential requirements for credit institutions
  and investment firms and amending Regulation (EU) No 648/2012 Text with EEA
  relevance}.
\newblock {\em European Commission\/}.

\bibitem[\protect\citeauthoryear{FCA}{FCA}{2014}]{fca:bipru}
FCA (2014).
\newblock {Prudential Sourcebook for Banks, Building Societies and Investment
  Firms (BIPRU)}.
\newblock Online; accessed 17 Feb 2014;
  \url{http://fshandbook.info/FS/html/handbook/BIPRU}.

\bibitem[\protect\citeauthoryear{Green and Kenyon}{Green and
  Kenyon}{2014}]{Green2014a}
Green, A. and C.~Kenyon (2014).
\newblock {Calculating the Funding Valuation Adjustment (FVA) of Value-at-Risk
  (VAR) based Initial Margin}.
\newblock {\em SSRN eLibrary\/}.
\newblock Available at http://ssrn.com/paper=2432281.

\bibitem[\protect\citeauthoryear{Hull and White}{Hull and
  White}{2014}]{Hull2014a}
Hull, J. and A.~White (2014, October).
\newblock {Risk Neutrality Stays}.
\newblock {\em {Risk}\/}~{\em 27(10)}.

\bibitem[\protect\citeauthoryear{Kenyon and Green}{Kenyon and
  Green}{2013}]{Kenyon2013d}
Kenyon, C. and A.~Green (2013).
\newblock {Pricing CDSs' capital relief}.
\newblock {\em {Risk}\/}~{\em 26\/}(10).

\bibitem[\protect\citeauthoryear{Kenyon and Green}{Kenyon and
  Green}{2014a}]{Kenyon2014b}
Kenyon, C. and A.~Green (2014a, September).
\newblock {Regulatory costs break risk neutrality}.
\newblock {\em {Risk}\/}~{\em 27(9)}.

\bibitem[\protect\citeauthoryear{Kenyon and Green}{Kenyon and
  Green}{2014b}]{Kenyon2014f}
Kenyon, C. and A.~Green (2014b, October).
\newblock {Regulatory Costs Remain}.
\newblock {\em {Risk}\/}~{\em 27(10)}.

\bibitem[\protect\citeauthoryear{Kenyon and Kenyon}{Kenyon and
  Kenyon}{2013}]{Kenyon2013a}
Kenyon, C. and R.~Kenyon (2013).
\newblock {DVA for Assets}.
\newblock {\em {Risk}\/}~{\em 26\/}(2).

\bibitem[\protect\citeauthoryear{Longstaff and Schwartz}{Longstaff and
  Schwartz}{2001}]{Longstaff2001a}
Longstaff, F. and E.~Schwartz (2001).
\newblock Valuing american options by simulation: A simple least-squares
  approach.
\newblock {\em The Review of Financial Studies\/}~{\em 14\/}(1), 113--147.

\bibitem[\protect\citeauthoryear{Reboul, Perrin, Morel, and Peace}{Reboul
  et~al.}{2014}]{Reboul2014a}
Reboul, P., B.~Perrin, G.~Morel, and J.~Peace (2014).
\newblock Corporate and investment banking outlook.
\newblock Technical report, Roland Berger and Nomura.

\end{thebibliography}

\end{document}